\def\prob{\mathop{\mathbb{P}}}
\def\ket#1{\left|#1\right\rangle}
\def\bra#1{\left\langle#1\right|}

\def\rank{\mathop{\rm rank}\nolimits}
\documentclass[aps,pra,twocolumn,notitlepage,
longbibliography]{revtex4-1}

\pdfoutput=1
\usepackage{xcolor}
\usepackage{graphics}
\graphicspath{{./circ/}}
\usepackage{hyperref}
\usepackage{amsfonts}
\usepackage{mathtools,amssymb,amsthm}

\usepackage[inline]{enumitem}
\setlist{nosep}

\usepackage{mathrsfs}

\begin{document}
\title{On maximum-likelihood decoding with circuit-level errors}
\author{Leonid P. Pryadko}

\affiliation{Department
  of Physics \& Astronomy, University of California, Riverside,
  California 92521, USA} 
\date{July 15, 2020}

\begin{abstract}
  Error probability distribution associated with a given Clifford
  measurement circuit is described exactly in terms of the circuit
  error-equivalence group, or the circuit subsystem code previously
  introduced by Bacon, Flammia, Harrow, and Shi.  This gives a
  prescription for maximum-likelihood decoding with a given
  measurement circuit.  Marginal distributions for subsets of circuit
  errors are also analyzed; these generate a family of related
  asymmetric LDPC codes of varying degeneracy.  More generally, such a
  family is associated with any quantum code.  Implications for
  decoding highly-degenerate quantum codes are discussed.
\end{abstract}

\maketitle

\section{Introduction}
Quantum computation offers exponential algorithmic speed-up for some
classically hard problems.  This promise is conditional in a
fundamental way upon the use of quantum error correction (QEC).
However, despite an enormous progress achieved both in theory and
experiment during the quarter century since the invention of
QEC\cite{shor-error-correct}, universal repetitive QEC protecting
against both phase and amplitude errors has not yet been demonstrated
in any of the qubit systems constructed to date.  This illustrates the
enormous difficulty of building quantum computers (QCs) with
sufficiently small error rates.

Given also the engineering difficulties with scaling up the number of
qubits\cite{Almudever-etal-2017}, it is important that on the
algorithmic side one tries to achieve every optimization possible.
Among other things, we would like to maximize the probability of
successful syndrome-based decoding with QEC.  Given the available
hardware, this requires choosing the optimal code, the optimal
measurement circuit, and the optimal decoder.  In particular, a
decoder should be designed for the specific syndrome measurement
circuit, as it must be aware of the associated correlations between
the errors\cite{Aliferis-Gottesman-Preskill-2006,%
  Wang-Fowler-Hollenberg-2011,Chubb-Flammia-2018}---at sufficiently
high error probabilities such correlations are present even in
fault-tolerant (FT) circuits designed to prevent a single fault from
affecting multiple data qubits.

In contrast, the standard approach is to use a decoder optimized for
the underlying code, regardless of the actual circuit used for
syndrome measurement.  In some cases, e.g., in the case of surface
codes with minimum-weight perfect matching (MWPM) decoder, some
leading-order correlations can be included in the edge weights of the
graph used for decoding\cite{Dennis-Kitaev-Landahl-Preskill-2002,%
  Wang-Fowler-Hollenberg-2011,Fowler-2012,%
  Fowler-Mariantoni-Martinis-Cleland-2012}.  However, such a scheme is
limited to codes and measurement circuits where MWPM can be done
efficiently, e.g., surface codes with single-ancilla measurement
circuits\cite{Fowler-Whiteside-McInnes-Rabbani-2012}, and certain
other classes or topological
codes\cite{Chamberland-Zhu-Yoder-Hertzberg-Cross-2020,%
  Chamberland-Kubica-Yoder-Zhu-2020}.  Further, there is necessarily a
decoding accuracy loss when measurement circuit is not simply
repeated, e.g., with code deformation or lattice
surgery\cite{Vuillot-etal-Terhal-2019}.

A feasible approach to building a decoder optimized for the specific
measurement circuit is to train a neural network (NN) using extensive
simulation data\cite{Torlai-Melko-2017,Krastanov-Jiang-2017,%
  Breuckmann-Ni-2017,%
  Jia-Zhang-etal-Guo-2019,Baireuther-etal-Beenakker-2018,%
  Chamberland-Ronagh-2019,Baireuther-etal-Beenakker-Obrien-2019,%
  Maskara-Kubica-JochymOConnor-2019}.
However, as is commonly the case with NNs, there is always a question
whether training has been sufficient to achieve optimal decoding
performance.

A parameterization of the probability distribution of correlated
quantum errors in terms of a spin model has been
recently considered by Chubb and Flammia\cite{Chubb-Flammia-2018}.  In
particular, they describe how such a formalism can be used for
maximum-likelihood (ML) decoding in the presence of circuit-level
noise.  However, Chubb and Flammia focused on larger circuits composed
of measurement blocks. Errors are assumed uncorrelated between the
blocks, while a model of error correlations at the output of each
block has to be constructed offline, e.g., using error propagation for
a Clifford measurement circuit.  In particular, Chubb and Flammia
stopped short of analyzing circuit-level errors, and only considered
numerically a toy model of correlated errors.

The goal of this work is to give an explicit numerically efficient
algorithm for analyzing error correlations resulting from a given
qubit-based Clifford measurement circuit, and for constructing
decoders optimized for such a circuit.  Main result is that such
correlations can be accounted for by using phenomenological error
model (no error correlations) with the circuit-associated subsystem
code constructed by Bacon, Flammia, Harrow, and
Shi\cite{Bacon-Flammia-Harrow-Shi-2015,Bacon-Flammia-Harrow-Shi-2017}.
Thus, any generic decoder capable of dealing with uncorrelated data
and syndrome measurement errors in highly-degenerate sparse-generator
subsystem codes can be rendered to account for circuit-level error
correlations.  Error correlations needed for implementing the scheme
of Chubb and Flammia are recovered by calculating the marginals of the
constructed distribution, which can be done in practice using Ising
model star-polygon transformations\cite{Strecka-2010}.  The
construction naturally includes additional correlations between errors
in different fault locations on the circuit.

An immediate application is for designing decoders approaching true ML
decoding for Clifford circuits, to be used in quantum memory.  In
particular, more accurate decoding and optimization, with the ability
to account for error rates' variations on individual qubits, could
help improve QEC to the level sufficient to pass the break-even point
and show coherent lifetimes longer than that of an unprotected qubit
in present-day or near-future devices.  Related techniques could also
be more widely applicable for design and analysis of FT protocols and
control schemes.  Examples include error correction with FT gadgets
like flag error correction\cite{Chamberland-Beverland-2017,%
  Chamberland-Cross-2018,Chao-Reichardt-2018}, schemes for protected
evolution, e.g., using code deformations where conventional approaches
may result in a reduced distance\cite{Vuillot-etal-Terhal-2019}, and
optimized single-shot error-correction
protocols\cite{Bombin-2015,Brown-Nickerson-Browne-2016,Campbell-2018}.

In addition, analysis of marginal error distributions and associated
families of asymmetric codes of varying degeneracy could become an
important tool in the theory of QECCs.  In particular, it could help
constructing better decoders for highly-degenerate quantum LDPC codes.
Also, thorough analytical understanding of error correlations could be
useful for fundamental analysis of thresholds, e.g., in order to
extend and/or improve the bounds in
Refs.~\onlinecite{Dumer-Kovalev-Pryadko-bnd-2015,%
  Kovalev-Prabhakar-Dumer-Pryadko-2018}.

Paper outline: After this Introduction, an overview of relevant
notations and results in quantum codes and multi-variable
Bernoulli distributions is given in Sec.~\ref{sec:background}.  Pauli
error channels with correlated errors, including circuit-level
correlations, are discussed in Sec.~\ref{sec:correlations}.
Corresponding marginal distributions are constructed in
Sec.~\ref{sec:marginal}, followed by a discussion of possible
implications for exact and approximate ML decoding with circuit-level
errors in Sec.~\ref{sec:decoding}.  Sec.~\ref{sec:discussion} gives
the concluding remarks.  

\section{Background}
\label{sec:background}
\subsection{Quantum codes}
Generally, an $n$-qubit quantum code $\mathcal{Q}$
is a subspace of the $n$-qubit Hilbert space
$\mathbb{H}_{2}^{\otimes n}$.  A quantum $[[n,k,d]]$ stabilizer code
is a $2^{k}$-dimensional subspace specified as a common $+1$
eigenspace of all operators in an Abelian \emph{stabilizer} group
${\cal S}\subset{\cal P}_{n}$, $-\openone\not\in{\cal S}$, where the
$n$-qubit Pauli group ${\cal P}_{n}$ is generated by tensor products
of single-qubit Pauli operators.  The stabilizer is typically
specified in terms of its generators,
${\cal S}=\left\langle S_{1},\ldots,S_{n-k}\right\rangle $. The weight
of a Pauli operator is the number of qubits that it affects. The
distance $d$ of a quantum code is the minimum weight of a Pauli
operator $E\in{\cal P}_n$ which commutes with all operators from the
stabilizer ${\cal S}$, but is not a part of the stabilizer,
$E\not\in{\cal S}$. Such operators act non-trivially in the code and
are called logical operators.

Subsystem codes\cite{Poulin-subs-2005,Bacon-subs-2006} are a
generalization of stabilizer codes where only some of the logical
qubits are used.  More precisely, given a stabilizer group
$\mathcal{S}$, the stabilized subspace $\mathcal{Q}_{\cal S}$ is
further decomposed into a tensor product of two subspaces,
$\mathcal{Q}_S=\mathcal{Q}_L\otimes \mathcal{Q}_G$, where
$\mathcal{Q}_L$ is a $2^k$-dimensional ``logical'' subspace used to
store the quantum information, while we do not care about the state of
the ``gauge'' qubits in the subspace $\mathcal{Q}_G$ after the
recovery.  Logical operators of the original code which act
non-trivially only in $\mathcal{Q}_G$, together with the operators
from the stabilizer group $\mathcal{S}$, generate the non-Abelian
gauge group $\mathcal{G}$ which fully characterizes the subsystem
code.  In particular, the center $Z(\mathcal{G})$ is formed by the
elements of the original stabilizer $\mathcal{S}$, up to a phase
$i^m$, $m\in\{0,1,2,3\}$.

A Pauli error $E$ that anticommutes with an element of the stabilizer,
$ES=-SE$, $S\in{\cal S}$, is called detectable.  Such an error results
in a non-zero syndrome ${\bf s}=\{s_1,\ldots,s_r\}$ whose bits
$s_i\in \{0,1\}$ are obtained by measuring the eigenvalues
$(-1)^{s_i}$ of the chosen set of stabilizer generators $S_i$,
$i=\{1,\ldots,r\}$, $r\ge n-k$.  Unlike in the case of classical
codes, there may be many equivalent (\emph{mutually degenerate})
errors resulting in the same syndrome.  For a subsystem code, errors
$E'$ degenerate with $E$, denoted $E'\simeq E$, have the form $E'=EG$,
where $G\in\mathcal{G}$ is an element of the gauge group; such errors
can not and need not be distinguished.  Non-trivial logical operators
of the subsystem code are logical operators of the original stabilizer
code which act non-trivially in ${\cal Q}_L$.  A \emph{bare} logical
operator $U$ acts trivially in ${\cal Q}_G$ and commutes with any
element of ${\cal G}$.  Such restrictions do not apply to
\emph{dressed} logical operators which are only required to commute
with elements of the stabilizer ${\cal S}$.  In any case,
multiplication by a logical operator $U$ gives an error with the same
syndrome but from a different equivalence class, $EU\not\simeq E$.

Analysis of error correction is conveniently done using quaternary, or an equivalent binary, representation of the Pauli
group\cite{gottesman-thesis,Calderbank-1997}.   A Pauli
operator $U\equiv i^{m}X^{\mathbf{u}}Z^{\mathbf{v}}$, where
$\mathbf{u},\mathbf{v}\in\{0,1\}^{\otimes n}$ and
$X^{\mathbf{u}}=X_{1}^{u_{1}}X_{2}^{u_{2}}\ldots X_{n}^{u_{n}}$,
$Z^{\mathbf{v}}=Z_{1}^{v_{1}}Z_{2}^{v_{2}}\ldots Z_{n}^{v_{n}}$, is
mapped, up to a phase, to a length-$2n$ binary vector
$\mathbf{e}=(\mathbf{u}|\mathbf{v})$.  Two Pauli operators $U_1$,
$U_2$ commute iff the symplectic inner product
\begin{equation}
  \label{eq:trace-product}
  \mathbf{e}_1\star \mathbf{e}_2^T
  \equiv \mathbf{e}_1\Sigma\,
  \mathbf{e}_2^T, \quad \Sigma\equiv\left(
    \begin{array}[c]{cc}
      & I_n\\ I_n&
    \end{array}\right),
\end{equation}
of the corresponding binary vectors is zero,
$\mathbf{e}_1\star\mathbf{e}_2^T=0$.  Here $I_n$ is the $n\times n$
identity matrix.  Thus, if $H=(H_Z|H_X)$ is an $m\times 2n$ binary
matrix whose rows represent stabilizer generators, and
$\widetilde{H}\equiv H\,\Sigma=(H_X|H_Z)$, then the syndrome $\mathbf{s}$
of an error $U_\mathbf{e}$ with binary representation
$\mathbf{e}=(\mathbf{u}|\mathbf{v})$ is given by
$\mathbf{s}^T={H}\mathbf{e}^T$.  If we similarly denote
$G=(G_X|G_Z)$ an $r\times 2n$ matrix formed by the gauge group
generators (for a stabilizer code, $G=\widetilde{H}$), the errors with binary
representation $\mathbf{e}$ and
$\mathbf{e}'=\mathbf{e}+\boldsymbol\alpha G$ are equivalent,
$\mathbf{e}'\simeq \mathbf{e}$, for any length-$r$ binary string
$\boldsymbol\alpha\in\mathbb{F}_2^{\otimes m}$.  Generally, 
$  G {H}^T=0$, and 
\begin{equation}
  \label{eq:rank}
 \rank H=n-k-\kappa, \quad \rank G=n-k+\kappa,
\end{equation}
where $\kappa$ is the number of gauge qubits.  Matrices $G$ and ${H}$
can be viewed as generator matrices of an auxiliary length-$2n$ CSS
code which encodes $2k$ qubits.  These correspond to a basis set of
$2k$ independent logical operators of the original subsystem code.  It
will be convenient to introduce a logical generating matrix $L$ such
that $LH^T=0$, $\rank L=2k$. Rows of $L$ map to basis logical
operators; a non-zero linear combination of the rows of $L$ must be
linearly independent from the rows of $G$.

Gauge or stabilizer generators can be measured with a Clifford circuit
which consists of ancillary qubit initialization and measurement in
the preferred (e.g., $Z$) basis, and a set of unitary Clifford gates,
e.g., single-qubit Hadamard $H$ and phase $P$ gates and two-qubit
CNOT.  Generally, a Clifford unitary $U$ maps Pauli operators to Pauli
operators, $E'=U^\dagger EU$, $E,E'\in\mathcal{P}_n$.  Ignoring the
overall phase (for complete description, see
Refs.~\onlinecite{Dehaene-DeMoor-2003,Aaronson-Gottesman-2004}), this
corresponds to a linear map of the corresponding binary vectors,
$(\mathbf{e}')^T=C \mathbf{e}^T$, where $C\equiv C_U$ is a symplectic
matrix with the property $C^T\Sigma C=\Sigma$.

\subsection{Bernoulli distribution}
\label{sec:Bernoulli}
Multi-variate Bernoulli distribution describes a joint probability
distribution of $m$ single-bit variables $x_i\in \mathbb{F}_2$, e.g.,
components of a binary vector ${\bf x}\in\mathbb{F}_2^m$.  Most
generally, such a distribution can be specified in terms of $2^m$
probabilities $p_{\bf x}\ge 0$, with normalization
$\sum_{{\bf x}\in\mathbb{F}_2^m}p_{\bf x}=1$.  A convenient
representation of such a distribution as an exponent of a polynomial
of $m$ binary variables with real coefficients was given by Dai et
al.\ in Ref.~\onlinecite{Dai-Ding-Wahba-2013}.  Namely, for a single
variable $x\in\{0,1\}$, we can write $\prob(x)=p_0^{1-x}p_1^x$ as a
product of two terms, where $p_0+p_1=1$ are the outcome probabilities.
For $m$ variables we have, similarly, the product of $2^m$ terms, 
\begin{eqnarray}
  \label{eq:Bernoulli}
  \prob(\mathbf{x})&=&p_{00\ldots 0}^{(1-x_1)(1-x_2)\ldots (1-x_m)}\nonumber \\
  & & \quad \times p_{00\ldots
    01}^{(1-x_1)(1-x_2)\ldots (1-x_{m-1})x_m}\ldots
  p_{11\ldots1}^{x_1x_2\ldots x_m},\quad
\end{eqnarray}
where the probabilities are assumed positive, $p_\mathbf{x}>0$. For
any given $\mathbf{x}\in\mathbb{F}_2^m$ only one exponent is non-zero
so that the result is $p_{\bf x}$.  Taking the logarithm and
expanding, one obtains the corresponding ``energy''
$\mathscr{E}\equiv -\ln \prob(\mathbf{x})$ as a polynomial of $m$
binary variables.  The corresponding coefficients can be viewed as
binary cumulants\cite{Dai-Ding-Wahba-2013}; presence of high-degree
terms indicates a complex probability distribution with highly
non-trivial correlations.

For applications it is more convenient to work with spin variables
$s_i=(-1)^{x_i}\equiv 1-2x_i\in\pm1$, and rewrite the energy function using
the general Ising representation first introduced by
Wegner\cite{Wegner-1971},
\begin{equation}
  \label{eq:wegner}
  \mathscr{E}\equiv \mathscr{E}(\{s_j\})=-\sum_b K_b R_b+{\rm const},\quad
  R_b=\prod_i s_i^{\theta_{ib}}, 
\end{equation}
parameterized by the binary spin-bond incidence matrix $\theta$ with
$m$ rows, and the bond coefficients $K_b$.  While most general
$m$-variate Bernoulli distribution requires $2^m-1$ bonds, in the
absence of high-order correlations significantly fewer terms may be
needed.  Of particular interest are distributions with sparse matrices
$\theta$, e.g., with bounded row and column weights, which also limits
the number of columns.  In the simplest case of independent
identically-distributed (i.i.d.) bits with equal set probabilities
$p_1=p$, $p_0=1-p$, we can take $\theta=I_m$, the identity matrix, and
all coefficients equal, $K=\ln[(1-p)/p]/2$.  The logarithm
$\ln[(1-p)/p]$ is commonly called a log-likelihood ratio (LLR); the
coefficient $K$ here and $K_b$ in Eq.~(\ref{eq:wegner}) are thus
called half-LLR coefficients.

\section{Error correlations in a Clifford measurement circuit}
\label{sec:correlations}
\subsection{Pauli error channel with correlations.}
\label{sec:Pauli-channel}
Consider the most general Pauli error channel
 \begin{equation}
  \label{eq:error-channel}
  \rho\to \sum\nolimits_{\mathbf{e}\in\mathbb{F}_2^{2n}} \prob(\mathbf{e}) E_\mathbf{e}\rho E_\mathbf{e}^\dagger,
\end{equation}
where $\prob(\mathbf{e})$ is the probability of an error
$E_\mathbf{e}$ with binary representation $\mathbf{e}$, with the
irrelevant phase disregarded.  Technically, $\prob(\mathbf{e})$
describes a $2n$-variate Bernoulli distribution.  The
probability $\prob(\mathbf{e})$ can be parameterized in terms
of a $2n\times m$ binary coupling matrix $\theta$ with $m<2^{2n}$
columns, and a set of coefficients $K_b$, $b\in\{1,\ldots,m\}$,
\begin{equation}
\prob(\mathbf{e};\theta,\{K_b\})=Z^{-1}\exp\left( \sum_b
  K_b(-1)^{[\mathbf{e}\,\theta]_b} \right),\label{eq:generic-prob}
\end{equation}
where $[\mathbf{e}\,\theta]_b$ in the exponent is the corresponding 
component of the row-vector $\mathbf{e}\,\theta$.  The normalization
constant $Z\equiv Z(\theta,\{K_b\})$ in Eq.~(\ref{eq:generic-prob}) is
the partition function of the Ising model in Wegner's form, cf.\ Eq.\
(\ref{eq:wegner}),
\begin{equation}
  \label{eq:Z}
  Z(\theta,\{K_b\})=\sum_{\{s_i\in \pm1\}}\prod_b e^{K_bR_b}.
\end{equation}
In the simplest case of independent $X$ and $Z$ errors, 
$\theta=I_{2n}$, the identity matrix, while $e^{2K_b}=(1-p_X)/p_X$ for
$b\le n$, and the corresponding expression with $p_X\to p_Z$ for
$n<b\le 2n$.  In the case of the depolarizing channel with error
probability $p$, we have, instead,
\begin{equation}
\theta=\left(
  \begin{array}[c]{c|c|c}
    I_n&0&I_n\\ \hline 0&I_n&I_n
  \end{array}\right), \quad e^{4K}=3(1-p)/p,\label{eq:theta-depolarizing}
\end{equation}
where the additional column block is to account for correlations between
$X$ and $Z$ errors.  Additional correlations between the errors can be
introduced by adding columns to matrix $\theta$ and the corresponding
coefficients $K_b$.

Given a probability distribution in the form (\ref{eq:generic-prob}),
it is easy to construct an expression for probability of an error
equivalent to $\mathbf{e}$ in a subsystem code with gauge generator
matrix $G$, extending the approach of
Refs.~\onlinecite{Dennis-Kitaev-Landahl-Preskill-2002,%
  Landahl-2011,Kovalev-Pryadko-SG-2015}, and reproducing some of the
results from Ref.~\onlinecite{Chubb-Flammia-2018}.  A substitution
${\bf e}\to {\bf e}+\boldsymbol\alpha G$ and a summation over
 $\boldsymbol\alpha$ gives
\begin{equation}
  \label{eq:corr-one}
  \mathbb{P}({\bf e}'\in\mathbb{F}_2^{2n}|{\bf e}'\simeq \mathbf{e})
  = 2^{\rank G-r}\,{Z\left(G\theta,\{K_b(-1)^{[{\bf e}\,\theta]_b}\}\right)
    \over Z\left(\theta,\{K_b\}\right)}.\;\;
\end{equation}
Here $G$ is an $r\times 2n$ binary matrix, cf.\ Eq.~(\ref{eq:rank}),
and the prefactor accounts for a possible redundancy in the summation.
Notice that the partition function in the numerator has the same
number of bonds $m$ as that in the denominator, but with the signs of
the coefficients $K_b$ corresponding to non-zero elements of the
binary vector $\boldsymbol\epsilon\equiv \mathbf{e}\,\theta$ flipped.
When both the gauge generator matrix $G$ and the error correlation
matrix $\theta$ are sparse, the incidence matrix $\Theta\equiv G\theta$ in
the numerator of Eq.~(\ref{eq:corr-one}) must also be sparse.

\subsection{Clifford measurement circuit and associated input/output
  codes.}
\label{sec:measurement}
\begin{figure}[htbp]
  \centering
  \includegraphics[width=0.4\columnwidth]{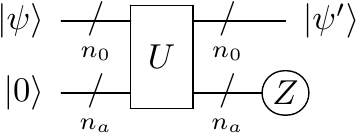}
  \caption{Generic circuit with $n_0$ data and $n_a$ ancillary qubits
    initialized to $\ket0$ and measured in $Z$ basis.  In practice,
    ancillary qubit can be measured during evolution, and subsequently
    reused after initialization.  However, there is no mechanism for
    adapting the gates to the measurement results.}
  \label{fig:meas-circ}
\end{figure}

Let us now consider error correlations resulting from a Clifford
measurement circuit.  Specifically, following
Refs.~\onlinecite{Bacon-Flammia-Harrow-Shi-2015,Bacon-Flammia-Harrow-Shi-2017},
consider an $n$-qubit circuit, $n=n_0+n_a$, with $n_0$ data qubits and
$n_a$ ancillary qubits.  First, ancillary qubits are initialized to
$\ket0$, second, a collection of Clifford gates forms a unitary $U$,
and finally the ancillary qubits are measured in the $Z$ basis, see
Fig.~\ref{fig:meas-circ}.  In the absence of errors and in the event
of all measurements returning $+1$ (zero syndrome), the corresponding
post-selected evolution is described by the matrix
\begin{equation}
  \label{eq:projected-evolution}
  V=(I^{\otimes n_0}\otimes \bra0^{\otimes n_a})\,U\,(I^{\otimes
    n_0}\otimes \ket0^{\otimes n_a}), 
\end{equation}
where $I$ is a single-qubit identity operator.  The circuit is assumed
to be a \emph{good error-detecting circuit} (good EDC), namely,
$V^\dagger V$ be proportional to the projector onto a subspace
$\mathcal{Q}_0\subseteq\mathbb{H}_2^{n_0}$,
\begin{equation}
V^\dagger V=c \Pi_{0},\quad c>0\label{eq:projector}
\end{equation}
see Def.~4 in Ref.~\onlinecite{Bacon-Flammia-Harrow-Shi-2017}.
Here ${\cal Q}_0$, called the \emph{input code}, is an $[[n_0,k,d_0]]$
stabilizer code encoding $k$ qubits with distance $d_0$.

A good EDC also defines an {\em output code}
$\mathcal{Q}_0'\subseteq \mathcal{H}_2^{\otimes n_0}$ which encodes
the same number of qubits $k$.  Indeed, since
$V^\dagger V=c\Pi_{0}$, matrix $V$ has only one non-zero
singular value, $\sqrt c$; this immediately gives $VV^\dagger=c\Pi_0'$,
with the projector onto a space
$\mathcal{Q}_0'\subseteq \mathcal{H}_2^{\otimes n_0}$ of the same
dimension $2^k$, the {output code}.  Moreover, for any input
state $\ket\psi\in\mathcal{Q}_0$, the output
$\ket\psi'\equiv V\ket\psi\in\mathcal{Q}_0'$ is in the output code,
and the corresponding transformation is a (scaled) Clifford unitary.

Even though the map between $\mathcal{Q}_0$ and $\mathcal{Q}_0'$ is
unitary, the distance $d_0'$ of the output code does not necessarily
equals $d_0$. In particular, adding a unitary decoding circuit on
output data qubits may be used to render $d_0'=1$.

\subsection{Errors in a Clifford circuit.}
\label{sec:errors-circuit}

Using standard circuit identities, any circuit
error $\mathcal{E}$ can be propagated forward to the output of the
circuit, thus giving an equivalent data error
${E}_0'(\mathcal{E})\in\mathcal{P}_{n_0}$ and the (gauge) syndrome
$\boldsymbol\sigma'(\mathcal{E})\in\mathbb{F}_2^{n_a}$ corresponding
to the measurement results.  Clearly, there is a big redundancy even
if phases are ignored, as many circuit errors can result in the same
or equivalent ${E}_0'(\mathcal{E})$ and
$\boldsymbol\sigma'(\mathcal{E})$.  The goal is to find the
conditional probability distribution for the equivalence class of the
output error ${E}_0'(\mathcal{E})$ given the measured value of
$\boldsymbol\sigma'(\mathcal{E})$.

\begin{figure}[htbp]
  \centering
  \includegraphics[width=0.6\columnwidth]{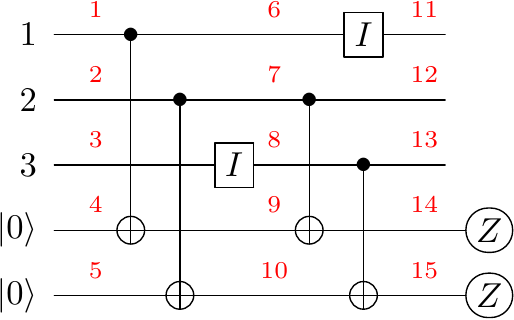}
  \caption{Circuit measuring generators $Z_1Z_2$ and $Z_2Z_3$ of a
    three-qubit toy code.  Digits indicate distinct circuit locations.
    Identity operators $I$ are inserted so that the number of circuit
    locations along each qubit wire be odd.}
  \label{fig:rep-circ}
\end{figure}

In a given Clifford circuit, consider $N$ possible \emph{error
  locations}, portions of horizontal wires starting and ending on a
gate or an input/output end of the wire.  For example, the circuit in
Fig.~\ref{fig:rep-circ} has $N=15$ error locations.  A Pauli error may
occur on any of these locations.  A \emph{circuit error} ${\cal E}$ is
a set of $N$ single-qubit Pauli operators without the phase,
$P_i\in \{I,X,Y,Z\}$, $i\in\{1,\ldots,N\}$.  When two circuit errors
are applied sequentially, the result is a circuit error whose elements
are pointwise products of Pauli operators with the phases dropped.
The algebra defined by such a product is isomorphic to the $N$-qubit
Pauli group without the phase.  This is an Abelian group which also
admits a representation in terms of length-$2N$ binary vectors
$\mathbf{e}\equiv (\mathbf{u}|\mathbf{v})\in \mathbb{F}_2^{2N}$.
Multiplication of two circuit errors amounts to addition of the
corresponding binary vectors $\mathbf{e}$.

With these definitions, error propagation through a Clifford circuit
can be described as products of the original circuit error ${\cal E}$
with trivial circuit errors which have no effect on the outcome.  The
collection of all such errors forms \emph{error equivalence group}
(EEG) of the circuit.  The generators of this group involved in error
propagation are trivial errors, each formed as a union of a
single-qubit Pauli $P_i$, $i\in\{1,\ldots,N\}$ (but not in the output
for data qubits, or right before the measurements for ancillary
qubits) with the result of error propagation of $P_i$ across the
subsequent gate.  For example, for the circuit in
Fig.~\ref{fig:rep-circ}, some of the EEG generators, with identity
operators dropped, are $\{Z_1,Z_6\}$, $\{X_4,X_9\}$,
$\{X_1,X_6,X_9\}$, and $\{Z_4,Z_6, Z_9\}$ (propagation across the
leftmost \textsc{CNOT}), and $\{X_3,X_8\}$, $\{Z_3,Z_8\}$ (propagation
across the leftmost identity gate labeled $I$).

In addition, for a circuit which includes qubits' initialization and
measurement, the list of EEG generators includes errors $Z_j$ on
ancillary qubits right after initialization to $\ket0$ and right
before the $Z$-basis measurement.  For the circuit in
Fig.~\ref{fig:rep-circ}, these are $\{Z_4\}$, $\{Z_5\}$ (ancillary
qubits right after initialization), and $\{Z_{14}\}$, $\{Z_{15}\}$
(ancillary qubits just before measurement).  It is important that
stabilizer group of the input code, after its elements are promoted as
circuit errors, forms a subgroup of thus constructed full circuit
EEG\cite{Bacon-Flammia-Harrow-Shi-2017}.  Same applies to the
stabilizer group of the output code.

In the 
binary representation, a single-qubit Hadamard gate connecting circuit
positions $i$ and $i'$ corresponds to a pair of generators with
non-zero elements $(u_i,v_i|u_{i'},v_{i'})=(10|01)$ and $(01|10)$
($X_i$ propagates into $Z_{i'}$ and v.v.), a phase gate similarly
corresponds to $(10|10)$ and $(01|11)$ ($Z_i$ propagates into $Z_{i'}$
and $X_i$ into $Y_{i'}$), and a CNOT gate $(10\,00|10\,00)$ (input
$Z_i$ on the control and an output $Z_{i'}$ on the same wire),
$(01\,00|01\,01)$ (input $X_i$ on the control and $X_{i'}$, $X_{j'}$
on the outputs), $(00\,01|00\,01)$ (target $X$ pass-through), and,
finally, $(00\,10|10\,10)$.  The generators for the single-qubit
trivial errors are even simpler, since a $Z_j$ maps to the weight-one
vector $(u_j,v_j)=(10)$.

For a given circuit error ${\cal E}$ with binary form
$\mathbf{e}\in\mathbb{F}_2^{2N}$, any equivalent error can be obtained
by adding a linear combination of thus constructed circuit EEG
generators $\mathbf{g}_j$.  It is convenient to combine the
corresponding generators into a generator matrix $G$ with $2N$
columns.  As discussed in the Appendix \ref{sec:app-circ}, for a good
EDC with the rank of the input-code stabilizer group $r_0=n_0-k$ and the
constant $c=1/2^\kappa$, see Eq.~(\ref{eq:projected-evolution}) and below,
the generator matrix has the rank
\begin{equation}
  \label{eq:rank-G}
  \rank G=2N-2n_0-f,\quad f\equiv n_a-\kappa-r_0\ge0.
\end{equation}
Generally, a non-zero value $f>0$ indicates a measurement redundancy.

With the circuit EEG generator matrix in place, it is easy to
construct a formal expression for the probability of an error in a
given equivalence class.  Namely, if the circuit error probability
distribution is given by an analog of Eq.~(\ref{eq:generic-prob}), the
probability of an error equivalent to a given
$\mathbf{e}\in\mathbb{F}_2^{2N}$ is proportional to the Ising partition
function,
\begin{equation}
  \mathbb{P}(\mathbf{e}'\in\mathbb{F}_2^{2N}| \mathbf{e}'\simeq \mathbf{e})={\rm Const}\,Z(G\theta,\{K_b(-1)^{[{\bf e}\theta]_b}\}),
  \label{eq:circuit-error-distr}
\end{equation}
cf.\ Eq.\ (\ref{eq:corr-one}).  It is also easy to find the
conditional probability distribution of output errors with a given
syndrome.  Namely, given the binary form of an output data error
$\mathbf{e}_0'\in\mathbb{F}_2^{\otimes 2n_0}$ and a syndrome vector
$\boldsymbol\sigma'\in \mathbb{F}_2^{\otimes n_a}$, we need to form
the corresponding vector
$\mathbf{e}\equiv
\mathbf{e}(\mathbf{e}_0',\boldsymbol\sigma')\in\mathbb{F}_0^{2N}$,
filling only the components corresponding to the output data qubits
and ancillary $X_j$ just before the measurements which give non-zero
syndrome bits in $\boldsymbol\sigma'$.

For ML decoding, we compare thus computed probabilities for all
inequivalent errors consistent with the measured syndrome
$\boldsymbol\sigma'$.  In particular, these include errors that differ
by a logical operator of the input (or, equivalently, the output)
code, since non-trivial logical operators are outside the circuit EEG.
In other words, length-$2N$ binary vectors corresponding to logical
operators are linearly independent from the rows of the circuit EEG
generator matrix $G$.  It is convenient to define a binary logical
generator matrix $L$ of dimension $2k\times 2N$, whose rows correspond
to mutually inequivalent logical operators of the input code.

For the ease of decoding, it is also convenient to introduce the
\emph{parity-check} matrix $H$, also with $2N$ columns, whose rows are
orthogonal to the rows of both matrices $G$ and $L$, and whose rank
satisfies
\begin{equation}
\rank H=2N-\rank G-\rank L=n_a-\kappa+r_0. \label{eq:rank-H}
\end{equation}
Clearly, $H$ is dual to a matrix combing the rows of $G$ and $L$.  As
in the case of a subsystem code, see Eq.~(\ref{eq:rank}), these
matrices can be seen as forming a half of a CSS code, with stabilizer
generator matrices $G_X=G$, $G_Z=H$, and $X$-logical operator
generator matrix $L_X=L$.

The orthogonality requirement for rows of $H$ can also be interpreted
in terms of the $N$-qubit Pauli group associated with the circuit.
Namely, the Pauli operators corresponding to the rows of the
symplectic dual matrix $\widetilde{H}=H\Sigma$, see
Eq.~(\ref{eq:trace-product}) and below, must commute with generators
of the circuit EEG.  This guarantees that each of these operators be a
\emph{spackle}, i.e., a circuit error where the single-qubit Pauli
operators in any time layer can be obtained by error propagation from
those in the previos time layer, see
Ref.~\onlinecite{Bacon-Flammia-Harrow-Shi-2017}.  Respectively, row
weights of $H$ scale linearly with the circuit depth.

\subsection{Circuit subsystem code}

The discussion in Ref.~\onlinecite{Bacon-Flammia-Harrow-Shi-2017}
focused on the special case of good EDCs where each qubit line is
\emph{required} to have an odd number of locations.  In this special
case, the circuit EEG can be seen as the gauge group of a quantum
subsystem code of length $N$ which encodes the same number of qubits
$k$ as the input/output codes, and has a distance not exceeding the
corresponding distances, $d\le \min(d_0,d_0')$.  Respectively, rows of
the matrix $\widetilde{H}$ which correspond to generators of the
stabilizer group of the subsystem code are necessarily given by linear
combinations of the rows of $G$. In addition, circuit errors
corresponding to the rows of the matrix $L$ can be seen as dressed
logical operators, while the bare logical operators which commute with
the elements of the circuit EEG can be constructed as spackles.

In practice, any circuit can be easily modified to satisfy this
additional requirement by inserting a null (identity) gate into each
qubit line with an even number of locations, see
Fig.~\ref{fig:rep-circ} for an example.  While it is not strictly
necessary to work with circuits that satisfy this requirement, it is
convenient, as the additional structure of the subsystem code can be
used to verify the validity of the constructed matrices.

However, once the matrices $G$, $H$, and $L$ are constructed, there is
no need to refer to the subsystem code.  In fact, the generator matrix
row-reduction transformation described in the following Section
\ref{sec:marginal} preserves the orthogonality and the relation
Eq.~(\ref{eq:rank-H}) between the ranks inherent in the CSS code map,
but not the structure of the circuit subsystem code.

\subsection{Circuit code distance.}
How good can a measurement protocol be?  What are the bounds on the
distance $d$ of the subsystem code associated with the circuit?

Generally, if $d_0$ and $d_0'$ are the distances of the input and the
output codes, the distance of the corresponding circuit code satisfies
$d\le \min(d_0,d_0')$.  This follows from the fact that a logical
operator of the input code, e.g., is naturally mapped to a (dressed)
logical operator of the circuit code.  An important result in
Ref.~\cite{Bacon-Flammia-Harrow-Shi-2017} is that one can always
design a fault-tolerant circuit so that the distance $d$ of the
corresponding subsystem code be as good as that of the input code,
$d=d_0$.

Unfortunately, circuit-code distance $d$ does not have a direct
relation to the probability distribution of the output errors; even
single-qubit output errors may remain undetected.  This is a well
known ``feature'' of quantum error-correcting codes operating in a
fault-tolerant regime, even for codes with single-shot
properties\cite{Bombin-2015,Brown-Nickerson-Browne-2016,Campbell-2018}.
Indeed, regardless of the circuit structure, errors on the data qubits
in the locations just before the circuit output will not be detected.

In comparison, with a formally defined circuit code, such an error can
be propagated back to the input layer and (when it is detectable,
e.g., if its weight is smaller than the distance $d_0'$ of the output
code) it would necessarily anticommute with one or more combination(s)
$Z_{\bf g}$ of the ancillary qubits.  The original error would thus be
detectable in the circuit code.  Causality does not permit such an
operation with actual circuit evolution.  Formally, this functionality
is removed due to assumed ancillary qubit initialization to $\ket0$.

Of course, even if a small-weight error goes undetected, it may get
corrected after one or more additional measurement rounds.  In
practice, when an error-correcting code is analyzed in a
fault-tolerant setting, the standard numerical procedure is to add a
layer of perfect stabilizer measurements (no measurement errors).
This guarantees that all small-weight errors at the end of the
simulation be detected, and thus recovers the distance $d>1$ of the
circuit code, without the need to violate causality.

\section{Marginal distributions for correlated errors}
\label{sec:marginal}
The circuit EEG fully describes correlations between the circuit
errors. However, it also contains a lot of excessive information: for
the purpose of error correction, we are only interested in the
distribution of the output errors and the syndrome, which are all
supported at the rightmost locations of the circuit.  In addition, the
large size and sparsity of the circuit generator matrix $G$ makes
decoding difficult, except with the simplest circuits.

Present goal is to reduce the number of independent variables, by
constructing the marginal distribution for a given subset of the
variables.  This amounts to a summation over the variables one is not
interested in, e.g.,%
\begin{equation}
  \label{eq:marginal-one}
  \mathbb{P}(e_{s+1},\ldots,e_m)
  =\sum_{e_1} \sum_{e_2}\ldots \sum_{e_s} \mathbb{P}(e_1,e_2,\ldots,e_m). 
\end{equation}
In the case of binary variables $e_i\in\{0,1\}$, both the original and
the resulting marginal distributions are multi-variate Bernoulli
distributions, and each can be described in
terms of the Ising energy function (\ref{eq:wegner}).
\subsection{Row-reduction transformation}
\label{sec:row-reduce}
\subsubsection{Generator matrix and the coupling coefficients}
\label{sec:row-reduce-G}

Given an $n$-variate Bernoulli distribution described by the coupling
matrix $\Theta$, e.g., $\Theta=G\theta$ in Eq.~(\ref{eq:corr-one}),
and a set of half-LLR coefficients $K_b$, $1\le b\le m$, what are the
corresponding parameters of the marginal distribution
(\ref{eq:marginal-one})?  In the equivalent Ising-model representation
(\ref{eq:wegner}), the goal is to describe the couplings between the
remaining spins after a partial summation.  Such a \emph{star-polygon}
transformation for a general Ising model has been constructed in
Ref.~\onlinecite{Strecka-2010}.  The transformation includes two
special cases long known in the Ising model literature: the Onsager's
star-triangle transformation\cite{Onsager-exact-1944} and the
(inverse) decoration transformation\cite{Naya-1954,Fisher-1959}.

It is convenient to derive the result directly, focusing on the
marginal distribution after a summation over just one spin variable
$s_i\in\pm1$ corresponding to $i$\,th row of $\Theta$.  Without
limiting generality, assume that the chosen row has $w$ non-zero
elements in positions $1$, $2$, \ldots, $w$, decompose the
corresponding bond variables $R_b=s_i T_b$, $T_b\in\pm1$,
$1\le b\le w$, and perform the summation explicitly (with the
additional one-half factor for convenience),%
\begin{equation}
  B_{\boldsymbol\tau}
  \!\equiv 
 {1\over2} \!\sum_{s_i=\pm1}\!\exp\Bigl({s_i \sum_{b=1}^w K_b T_b}\Bigr)=
  \cosh\Bigl(\sum_{b=1}^w K_b T_b\Bigr),%
  \label{eq:cosh}
\end{equation}
where $\boldsymbol \tau\in\mathbb{F}_2^w$ is a composite index with
elements $\tau_b$ such that $T_{b}=(-1)^{\tau_b}$.  To exponentiate
this expression, rewrite $B_{\boldsymbol\tau}$ by analogy with
Eq.~(\ref{eq:Bernoulli}),
\begin{eqnarray}
  \nonumber
  B_{\boldsymbol\tau}
  &=&
      B _{00\ldots0}^{{1+\tau_1\over2}{1+\tau_2\over2}\ldots {1+\tau_w\over2}}\,
      B_{00\ldots01}^{{1+\tau_1\over2}{1+\tau_2\over2}\ldots
      {1+\tau_{w-1}\over2}{1-\tau_w\over2}} \\ 
  & & 
     \qquad\qquad\;\;\;\times\ldots\; B_{11\ldots1}^{{1-\tau_1\over2}{1-\tau_2\over2}\ldots {1-\tau_w\over2}},
            \label{eq:poly-ising}
\end{eqnarray}
where the coefficients $B_{\ldots}$ in the base of the exponents are
the hyperbolic cosines (\ref{eq:cosh}) of the sum of coefficients
$\pm K_b$ with the signs fixed, and matching exactly the signs in the
exponents.  As in Eq.~(\ref{eq:Bernoulli}), after a substitution of
any given $\boldsymbol\tau\in \mathbb{F}_2^w$, only one term with the
correct index $\boldsymbol\tau$ remains in the product.  The modified
bonds and the corresponding coefficients $K_b'$ are obtained by
expanding the polynomial in the exponent of Eq.~(\ref{eq:poly-ising}).
Because of the symmetry of the hyperbolic cosine, only even-weight
products of the original bonds result from this expansion.  Thus, in
general, for an original row of weight $w$, the corresponding $w$
columns are combined to produce $w'=2^{w-1}-1$ even-weight column
combinations, a change of $\Delta w=2^{w-1}-w-1$ columns.

Specifically, for a row of weight $w=1$, the transformation amounts to
simply dropping the row and the corresponding column of $\Theta$.  The
values of $K_b$ remain the same, except for the one value that is
dropped.

For a row of weight $w=2$, only the sum of the corresponding columns
is retained in $\Theta$, with the coefficient
$$
K_{1,2}'={1\over 2}\ln {\cosh (K_1+K_2)\over \cosh (K_1-K_2)}\equiv
{1\over 2} \ln{B_{00}\over B_{01}},
$$
cf.\ Eq.~(\ref{eq:cosh}).  Equivalently, 
$\tanh K_{1,2}'=\tanh K_1 \tanh K_2$.

For a row of weigth $w=3$, the three columns of the original matrix
$\Theta$ are replaced by their pairwise sums, with the coefficient
$$
K_{1,2}'={1\over 4}\ln {B_{000}B_{001}\over B_{010}B_{011}}
$$
for the combination of the first two columns.  The remaining
coefficients $K_{2,3}'$ and $K_{3,1}'$ can be obtained with cyclic
permutations of the indices.

For a row of weight $w=4$, the four columns of the original matrix $\Theta$
are replaced with six pairwise sums and the seventh column combining
all four original columns, with the coefficients, e.g.,
\begin{eqnarray*}
  K_{1,2}'&=&{1\over 8}\ln {B_{0000}B_{0001}B_{0010}B_{0011}\over
              B_{0100}B_{0101}B_{0110}B_{0111}},\\
  K_{1,2,3,4}'&=&{1\over 8}\ln {B_{0000}B_{0011}B_{0101}B_{0110}\over
              B_{0001}B_{0010}B_{0100}B_{0111}}.
\end{eqnarray*}

In general, the coefficient $K'_J$ in front of the product of $T_b$
with indices $b$ in an (even) subset $J\subseteq \{1,2,\ldots,w\}$ is
given by the sum of logarithms of the hyperbolic cosines
$B_{\boldsymbol\tau}$ with $\tau_1=0$ (this accounts for symmetry of
hyperbolic cosine), with the coefficients $\pm1/2^{w-1}$, where the
sign is determined by the parity of the weight of the subset
$\boldsymbol{\tau}[J]$ restricted to $J$.  It is easy to check that
the numbers of positive and negative coefficients always match.
Respectively, the coefficients for high-weight products are typically
small in magnitude.

\subsubsection{Transformation for a parity check matrix}
\label{sec:row-reduce-H}

The row-reduction transformation can also be written in terms of the
\emph{parity-check} matrix ${\rm H}$, also with $m$ columns, and
\emph{dual} to $\Theta$, such that
\begin{equation}
  \label{eq:matrix-dual}
  {\rm H}\Theta^T=0 \text{\ \ \;and\ \ }  \rank {\rm H}=m-\rank \Theta.
\end{equation}
To this end, consider the row-reduction of $\Theta$ as a combination of
the following elementary column steps:
\begin{enumerate}[label=(\roman*)]
\item The 1st column of $\Theta$ is added to columns $2$, $3$, \ldots, $w$
  of $\Theta$; as a result the $i$\,th row of $\Theta$ has a non-zero element only
  in the first position.
\item The 1st column of thus modified $\Theta$ is dropped, which leaves the
  $i$\,th row zero---it may be dropped as well;
\item If $w>2$, $2^{w-1}-w$ combinations of two, three, \ldots, $w-1$
  columns with indices $b\le w-1$ are added to the matrix $\Theta$.  These
  can be sorted by weight so that each added column be a combination
  of exactly two existing columns in the modified $\Theta$.
\end{enumerate}
 The corresponding steps for ${\rm H}$
are:
\begin{enumerate}[label=(\roman*$'$)]
\item Columns $b\in \{2,3,\ldots,w\}$ of ${\rm H}$ are added to its 1st
  column, which becomes identically zero as a result.
\item Drop all-zero 1st column from thus modified ${\rm H}$.
\item For each column, e.g., $b'$, added to $\Theta$ as a linear
  combination of two existing columns $b_1$ and $b_2$, ${\rm H}$ acquires a
  new row with the support on $\{b_1,b_2,b'\}$  to express this
  linear relation.
\end{enumerate}
It is easy to check that row orthogonality, ${\rm H}\Theta^T=0$, is
preserved.  Also, the rank of $\Theta$ is reduced by one, while the
increase of the rank of ${\rm H}$ matches the number of columns added
in step (iii$'$), so that the exact duality (\ref{eq:matrix-dual}) is
preserved.

\subsection{Marginal distribution for error equivalence classes}
\label{sec:marginal-equivalence}
This analysis is easily carried over to the problem of syndrome-based
ML decoding for an $[[n,k]]$ subsystem code under a Pauli channel
characterized by a $2n\times m$ matrix $\theta$ and a set of $m$
half-LLR coefficients $\{K_b\}$, see Sec.~\ref{sec:Pauli-channel}.
Given a gauge generator matrix $G$, the probability of an error
equivalent to $\mathbf{e}$ is given by Eq.~(\ref{eq:corr-one}).
Generally, for ML decoding we need to choose the largest of the
$2^{2k}$ partition functions
\begin{equation}
  Z(G\theta,\{K_b(-1)^{[\mathbf{e}'\theta]_b}), \quad 
  \mathbf{e}'=\mathbf{e}+\boldsymbol\alpha L,\quad
  \boldsymbol\alpha\in\mathbb{F}_2^{2k}.\label{eq:max-choice}
\end{equation}
Typically, this needs
to be done for a large number of error vectors $\mathbf{e}$.  Can the
calculation be simplified en masse by doing partial summation over the
spins corresponding to \emph{all} rows of $G\theta$ as described in
the previous section?

The structure of the logical operators can be accounted for by
extending the rows of the generator matrix which now has two row blocks,
\begin{equation}
  \label{eq:matr}
  \Theta=\left(
    \begin{array}[c]{c}      G\theta    \\ \hline L\theta
    \end{array}\right),\quad \tilde{K}_b=K_b (-1)^{[{\bf e}\theta]_b},
\end{equation}
and the half-LLR coefficients $\tilde{K}_b$,
$b\in\{1, 2, \ldots, m\}$.  A matching parity check matrix ${\rm H}$
is dual to $\Theta$, see 
Eq.~(\ref{eq:matrix-dual}).

\subsubsection{Independent $X$ and $Z$ errors}

Let us first consider the simpler case of independent $X$ and $Z$
errors, where matrix $\theta=I_{2n}$.  In this case
${\rm H}=H= \widetilde{H}\Sigma$, where $\widetilde H$ is a generator
matrix of the code's stabilizer group, see Eq.~(\ref{eq:rank}).
Marginal distribution being independent of the choice of the generator
matrix, use row transformations and column permutations to render%
\begin{eqnarray}\label{eq:row-reduced-G}
  \Theta
  &=&
      \left(
      \begin{array}[c]{c}G\\ \hline L    
      \end{array}\right)= 
  \left(
  \begin{array}[c]{c|c|c}I_r&A&B\\ \hline &I_{2k}&C    
  \end{array}\right),\\
  {H}&=&
  \left(
  \begin{array}[c]{c|c|c}B^T+C^TA^T& C^T&I_\ell
    \end{array}\right),\label{eq:row-reduced-H}
\end{eqnarray}
where 
matrices $B$ and $C$ have $\ell\equiv 2n-r-2k$ columns, and
$r= \rank (G)$.  The matrices $\Theta$ and ${H}$ are
mutually dual as can be immediately verified.

Row-reduction operations applied to each row in the upper block of
$\Theta$ correspond to:
\begin{enumerate}[label=(\roman*$''$)]
\item Column operations to set both blocks $A$ and $B$ to zero, and
  conjugate column operations on ${H}$ to set its left-most
  column block to zero.
\item Drop the upper row-block of the obtained $\Theta$, as well as
  the left-most column blocks of the resulting $\Theta$ and ${H}$.
\item Add an extra column block $M_1+CM_2$ to the resulting $\Theta$,
  where columns of $M_1$ and $M_2$ specify the linear combinations of
  the columns in its two remaining blocks, and a matching row-block to
  ${H}$.
\end{enumerate}
As a result, the transformed matrices acquire the form 
\begin{eqnarray}
  \label{eq:row-transformed-G}
  \Theta'
  &=&\left(
      \begin{array}[c]{c|c||c} I_{2k}&C&M_1+CM_2 \end{array}\right),\\
  \label{eq:row-transformed-H}
  {H}'
  &=&\left(
      \begin{array}[c]{c|c||c} C^T&I_\ell& 0\\ \hline
        M_1^T& M_2^T&I
      \end{array}\right),
\end{eqnarray}
where double vertical lines are used to separate the newly added columns. 

When the described transformation is applied to any of the original
partition functions (\ref{eq:max-choice}), the result is just an
exponential $\exp\left(\sum_b K_b^{\rm (fin)}\right)$ of the sum of
the final half-LLR coefficients.  Can identical columns of the final
generator matrix (\ref{eq:row-transformed-G}) be similarly combined to
simplify the structure of the final marginal distribution for error
equivalence classes?  The answer is yes, as long as we account for the
effect of the error ${\bf e}$ on the values of the coefficients
$K_b^{\rm (fin)}$.

In fact, it is easy to check that the row-reduction transformations in
Sec.~\ref{sec:row-reduce} are such that the additional signs in
Eq.~(\ref{eq:max-choice}) only affect the signs of the coefficients
$K_b^{\rm (fin)}$.  Moreover, these signs correspond to the bits of
the transformed error vector, cf.\ Eq.~(\ref{eq:row-transformed-G}),
\begin{equation}
  \label{eq:transformed-e}
  {\bf e}^{\rm (fin)}=\left(
    \begin{array}[c]{c|c||c}
      {\boldsymbol\varepsilon}_1&{\boldsymbol\varepsilon}_2
      &{\boldsymbol\varepsilon}_1 M_1 +{\boldsymbol\varepsilon}_2 M_2
    \end{array}\right),
\end{equation}
where the vector ${\boldsymbol\varepsilon}_1$ selects the equivalence
class, and
${\boldsymbol\varepsilon}_2={\bf s}+{\boldsymbol\varepsilon}_1 C$,
with $\mathbf{s}\equiv {\bf e} {H}^T$ the original syndrome.  Clearly,
the right-most blocks in Eqs.~(\ref{eq:row-transformed-G}) and
(\ref{eq:transformed-e}) are obtained from
$\boldsymbol\varepsilon\equiv
(\boldsymbol\varepsilon_1|\boldsymbol\varepsilon_2)$ with
$2k+\ell=2n-r$ components as a right product with the combined matrix
$M={M_1\choose M_2}$.  All $\ell_0\equiv 2n-r$ components of
$\boldsymbol\varepsilon$ being independent, there are
\begin{equation}
  \label{eq:count-cols-H-fin}
  m'_{\rm max}=2^{\ell_0}-1
\end{equation}
non-trivial combinations.  Combining identical columns in the
transformation from $\boldsymbol\varepsilon$ to ${\bf e}^{\rm (fin)}$,
we can ensure that the final matrices contain no more than
$m'_\mathrm{max}$ columns. 

With Eq.~(\ref{eq:rank}), $r=n-k+\kappa$, so that $\ell_0=n+k-\kappa$,
where $\kappa$ is the number of gauge qubits in the subsystem code.
For a stabilizer code, $\kappa=0$, thus $\ell_0=n+k$.  Clearly, the
latter is just the number of logical generators, $2k$, plus the number
of independent syndrome bits, $n-k$.

\subsubsection{A more general Pauli channel}
Instead of considering most general situation, consider an important
case of a Pauli channel where any single-qubit error has a non-zero
probability.  Then, the incidence matrix can be chosen to have an
identity-matrix block, $\theta=(I_{2n}|T)$, where $T$ is an
$(m-2n)\times 2n$ binary matrix.  As a result, the matrix $\Theta$ and
the half-LLR coefficients in Eq.~(\ref{eq:matr}) both acquire
additional blocks of linearly-dependent components, while the
parity-check matrix dual to $\Theta$ can be chosen in the form
\begin{equation}
  \label{eq:correlated-H}
  {\rm H}=\left(
    \begin{array}[c]{c|c}
      H&0\\ \hline T^T&I_{m-2n}
    \end{array}
\right).
\end{equation}
Since the relevant error in Eq.~(\ref{eq:matr}) is
$\mathbf{e}\theta=(\mathbf{e}|\mathbf{e}T)$, the lower row-block of
${\rm H}$ gives a zero syndrome, just like the lower row-block in
Eq.~(\ref{eq:row-transformed-H}).  Basically, after degeneracy is
taken into account, the number of independent components of
$\mathbf{e}\theta$ is $2n-r$, the same as for $\bf e$; final matrices
$\Theta'$ and ${\rm H}'$ can always be constructed to have
$m'\le m'_\mathrm{max}$ components given by
Eq.~(\ref{eq:count-cols-H-fin}).  Of course, the actual resulting
matrices, as well as the final half-LLR coefficients do depend on the
assumed error model.

How much freedom is there to choose the matrices $\Theta'$ and
${\rm H}'$?  For the purpose of ML decoding, we need to go over the
entire linear space $\mathcal{C}_{\Theta'}$ generated by the rows of
$\Theta'$; the choice of basis is irrelevant.  The same is true
regarding the parity check matrix ${\rm H}'$.  These are the same
symmetries as for a generator and a parity-check matrices of a binary
code.

In essence, the original quantum subsystem code with gauge $G$ and
logical $L$ generator matrices has been transformed into a
\emph{classical} binary code, with the transformation dependent in a
non-trivial fashion on the error probability distribution.  This
binary code has length $m'\le m'_\mathrm{max}$, and it encodes $2k$
bits.

Further, any non-trivial linear combination of rows of $L\theta$ with
rows of $G\theta$ has weight lower-bounded by the distance of the
quantum code, which gives the lower bound $d'\ge d$ on the distance of
this binary code.  In general, given the structure of the row-reduction
transformation, the distance may be quite a bit larger, possibly
scaling linearly with $m'$.  Notice, however, that with highly
non-uniform error probability distribution, more relevant parameter is
not the distance, but the corresponding quantity weighted with
half-LLR coefficients $K_b^{\rm(fin)}$, related to the probability of
the most-likely logical error.  By construction, this quantity is
exactly the same as for the original quantum code.

Let us consider an important case of a sparse original parity check
matrix ${\rm H}$, e.g., with row and column weights bounded.  This
requires a low-density parity-check (LDPC) code with a sparse
stabilizer generator matrix $\widetilde H= H\Sigma$, and an error
channel with the generator matrix $\theta=(I_{2n}|T)$ also sparse.
When acting on the parity check matrix, each row-reduction
transformation drops a column of ${\rm H}$, and may also add one or
more rows of weight $3$, see Sec.~\ref{sec:row-reduce-H}.  Thus, the
parity-check matrix of the classical binary code describing the
marginal probability distribution of error equivalence classes must
also be sparse, with row weights not exceeding $w'\le \max(w,3)$,
where $w$ is the maximum row weight of ${\rm H}$ in
Eq.~(\ref{eq:correlated-H}).

\subsection{Marginal distribution for output errors in a good
  measurement circuit}
\label{sec:marginal-output}

This discussion also applies to the code associated with the error
equivalence group of a good EDC.  In this case the matrices $G$
and $H$ have $2N$ columns each, where $N$ is the number of circuit
locations, and their ranks are given by Eqs.~(\ref{eq:rank-G}) and
(\ref{eq:rank-H}).  Just as any circuit error can be pushed all the
way to the right, row-reduction can also be done starting with the
bits at the beginning of the circuit and pushing toward its output.
This way, a circuit error equivalence class can be characterized by
\begin{equation}
\ell_1=2N-\rank G=2n_0+f = (n_0+k)+(n_a-\kappa)\label{eq:ell-one}
\end{equation}
bits, where $n_0+k$ is the number of linearly-independent error
equivalence classes in an $[[n_0,k]]$ stabilizer code and $n_a-\kappa$
is the number of syndrome bits measured in the circuit.
Alternatively, as in a subsystem code with $\kappa$ gauge qubits,
$\ell_0=n_0+k-\kappa$ is the exponent in
Eq.~(\ref{eq:count-cols-H-fin}), and $n_a$ is the number of additional
error positions right before the measurements.  As in the previous
section, this gives an upper bound on the maximum number $M'$ of
columns in the matrices $\Theta'$ and ${\rm H}'$ that may be
necessary,%
\begin{equation}
  \label{eq:m-bound}
  M'\le M'_{\rm max}=2^{\ell_1}-1.
\end{equation}
After row-reduction for all generators of the circuit EEG, we get a
classical $[M',2k,d']$ binary code, where $k$ is the number of encoded
qubits, and $d'\ge d$.

Is this an LDPC code?  This question has not been answered in the
previous section: the parity check matrix ${H}$ of the circuit code is
not necessarily sparse as its row weights scale linearly with circuit
depth.
To analyze the sparsity of the output-error parity-check matrix
${\rm H}'$, write it in a block form similar to
Eq.~(\ref{eq:correlated-H}),
\begin{equation}
  \label{eq:block-H}
 {\rm H'}=\left(
    \begin{array}[c]{c|c}
     {\rm H}_0'& 0\\ \hline       H_0''& H_1''
    \end{array}
  \right),
\end{equation}
where the upper row-block contains only the original columns of the
circuit EEG parity check matrix ${\rm H}$ remaining after
row-reduction steps (i$'$) and (ii$'$), while any row with exactly
three non-zero entries added in a step (iii$'$) goes to the lower
row-block.  Further, if we assume circuit EEG ${\rm H}$ in the form
(\ref{eq:correlated-H}), with the lower row-block of bounded weight
(e.g., $w=3$ for depolarizing errors), any potentially large-weight
row in ${\rm H}_0'$ must correspond to (i) a stabilizer generator of
the output code, or (ii) a generator of the group $\mathcal{H}'$, see
Ref.~\onlinecite{Bacon-Flammia-Harrow-Shi-2017}.  All of these have
bounded weights for any bounded-weight stabilizer LDPC code with a
measurement circuit where stabilizer generators are measured
independently and with a bounded number of ancillary qubits.  On the
other hand, these conditions do not generally hold for any family of
concatenated codes based on a given code, or for subsystem codes where
a stabilizer generator may have an unbounded weight or cannot be
expressed as a product of gauge generators with a bounded number of
terms.  Nevertheless, even in these cases, given a potentially very
large number of columns (\ref{eq:m-bound}), it is reasonable to expect
the final ${\rm H}'$ to be a sparse matrix, with only a small fraction
of non-zero elements.

\section{Decoding strategies}
\label{sec:decoding}
\subsection{Decoding based on circuit EEG}
The present approach is to analyze error propagation in a Clifford
measurement circuit in terms of its circuit EEG characterized by a
logical generator matrix $L$ and either a generator matrix $G$ or a
parity check matrix $H$.  With a minor constraint on the circuit,
these correspond to the circuit subsystem code constructed in
Refs.~\onlinecite{Bacon-Flammia-Harrow-Shi-2015,Bacon-Flammia-Harrow-Shi-2017}.
More generally, these matrices form (a half of) a CSS code, with
generator matrices $G_X=G$ and $G_Z=H$, while rows of $L$ define
$X$-type logical operators.  Simply put, any decoder capable of
dealing with a CSS code with uncorrelated $X$-type errors can now be
used to account for error correlations in a given measurement circuit.

Additional correlations can also be accounted for.  Assuming a Pauli
error channel (with or without correlations between different circuit
locations) characterized by a coupling matrix $\theta$ and a set of
half-LLR coefficients $K_b$ (one per column), probability of a circuit
error equivalent to a given one is given by the Ising partition
function (\ref{eq:circuit-error-distr}), with the spin-bond coupling
matrix $G\theta$.

As already discussed, for ML decoding we need to find the largest of
the Ising partition functions (\ref{eq:max-choice}).  Such a
calculation can be expensive.  Indeed, given a code encoding $k$
qubits, we need to compute and choose the largest of all $2^{2k}$
partition functions corresponding to all non-trivial sectors; this can
only be done in reasonable time for a code with $k$ small.
Previously, a computationally efficient ML decoder using this approach
and 2D tensor network contraction for computing the partition
functions has been constructed for surface
codes\cite{Bravyi-Suchara-Vargo-2014} in the channel model (perfect
syndrome measurement).

Feasible approaches for evaluating the partition functions
(\ref{eq:max-choice}) include tensor network contraction (see, e.g.,
in Ref.~\onlinecite{Hauru-etal-tensor-2018}, for a 3D network
contraction with complexity scaling as $\propto n\chi^9$, where $\chi$
is the bond dimension) and Monte-Carlo (MC) methods constructed
specifically for efficient calculations of free energy differences,
e.g., the non-equilibrium dynamics method \cite{deKoning-etal-2000} or
the classical Bennett acceptance ratio\cite{Bennett-1976}.  In
application to surface codes in FT regime, MC calculations are in
essence simulations of bond-disordered 3D Ising model; such
calculations can be done using GPU\cite{Preis-etal-2009},
FPGA\cite{Gilman-etal-2016}, or TPU\cite{Yang-etal-2019} hardware
acceleration.

Notice also that the circuit code is extremely degenerate: with
Hadamard, Phase, and CNOT gates the rows of the generator matrix $G$
have (quaternary) weights not exceeding three.  On the other hand, the
row weights of the parity-check matrix $H$ scale linearly with the
circuit depth.  By these reasons, iterative decoders like belief
propagation (BP) are expected to fare even worse than with the usual
(not so degenerate) quantum LDPC codes\cite{Poulin-Chung-2008,%
  Liu-Poulin-2019,Rigby-Olivier-Jarvis-2019}.

\subsection{Generator-based decoding via marginal distributions}
\label{sec:dec-marg-G}

The calculation of the Ising partition functions needed for ML
decoding can be simplified via partial summation over a
subset of the spins.  Technically, this corresponds to using a
marginal distribution for the subset of variables needed, as discussed
in Sec.~\ref{sec:marginal}.  

Denote $\mathcal{V}$ the set of rows of the original generator matrix
$G$ [also, rows in the upper block of $\Theta$ in
Eq.~(\ref{eq:matr})].  Then, row-reduction in an increasing sequence
of subsets%
\begin{equation}
{\cal I}_0\equiv \emptyset\subset {\cal I}_1\subset {\cal I}_2\subset
\ldots \subset {\cal I}_s\equiv \mathcal{V}\label{eq:subset-sequence}
\end{equation}
defines a sequence of mutually-dual pairs of matrices
$\{\Theta^{(j)},{\rm H}^{(j)}\}$ with $m_j$ columns, where
\begin{equation}
  \label{eq:matr-seq}
  \Theta^{(j)}=\left(
    \begin{array}[c]{c}
      G^{(j)}\\ \hline L^{(j)}
    \end{array}\right),\quad j\in\{0,\ldots,s\}.
\end{equation}
The ranks of logical generator matrices remain the same,
$\rank L^{(j)}= 2k$, while the sequence $r_j\equiv \rank G^{(j)}$ is
decreasing with increasing $j$, ending at $r_s=0$.  In essence, this
defines a sequence of asymmetric (half) CSS codes $[[m_j,2k]]$, with
generator matrices $G_X^{(j)}=G^{(j)}$ and $G_Z^{(j)}={\rm H}^{(j)}$,
where rows of $L^{(j)}$ define $X$-type logical operators.  The
sequence ends with a CSS code with an empty $X$-generator matrix,
i.e., a classical binary code with the parity check matrix
${\rm H}^{(s)}$.

For each of the codes in the sequence (\ref{eq:matr-seq}), there is
also a set of half-LLR coefficients $\{ K_b^{(j)},1\le b\le m_j\}$.
Given an error vector ${\bf e}\in \mathbb{F}_{2}^{m_j}$ matching the
syndrome, ML decoding can be done by  choosing the
largest of the $2^{2k}$ Ising partition functions [cf.\
Eq.~(\ref{eq:max-choice})]
\begin{equation}
Z\left(G^{(j)},\{K_b^{(j)} (-1)^{e_b'}\}\right),\quad {\bf e'}={\bf
  e}+\boldsymbol\alpha L^{(j)},\quad \boldsymbol\alpha\in
\mathbb{F}_2^{2k}.\label{eq:Z-max-reduced}
\end{equation}
By construction, the result should be the same for every $j\le s$.
However, the complexities for computing the partition functions may
differ dramatically.

One possibility for exact ML decoding is thus to choose a subset
$\mathcal{I}\subset \mathcal{V}$ to optimize the partition function
calculation.  In particular, one option is to choose $\mathcal{I}_1$
such that all rows of weights one, two, and three are eliminated from
the corresponding matrix $G^{(1)}$.  This minimizes the number of
columns in the exact generator matrix of the marginal-distribution.
Even though all rows in the original circuit EEG generator matrix $G$
have row weights not exceeding three, row-reduction on rows of weight
three tends to create higher-weight rows.  Thus, in general, $r_1>0$:
the resulting partition function is not expected to be trivial.

\begin{figure}[htbp]
  \centering
  \includegraphics[width=\columnwidth]{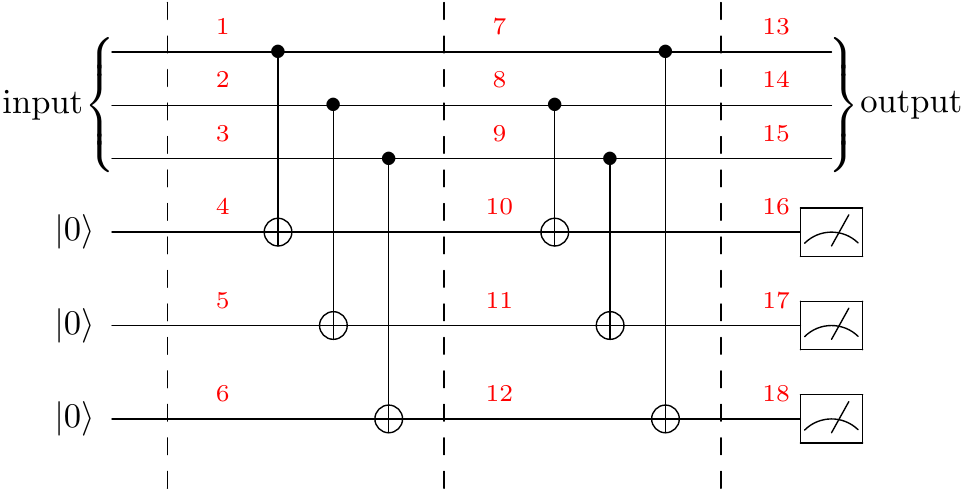}
  \caption{Single-cycle circuit measuring the overcomplete set
    $\{Z_1Z_2,Z_2Z_3,Z_3Z_1\}$ of stabilizer generators for the toy
    code with $n_0=3$ data and $n_a=3$ ancillary qubits.}
  \label{fig:3-3-1-circ}
\end{figure}

\begin{figure}[htbp]
  \centering
  \includegraphics[width=\columnwidth]{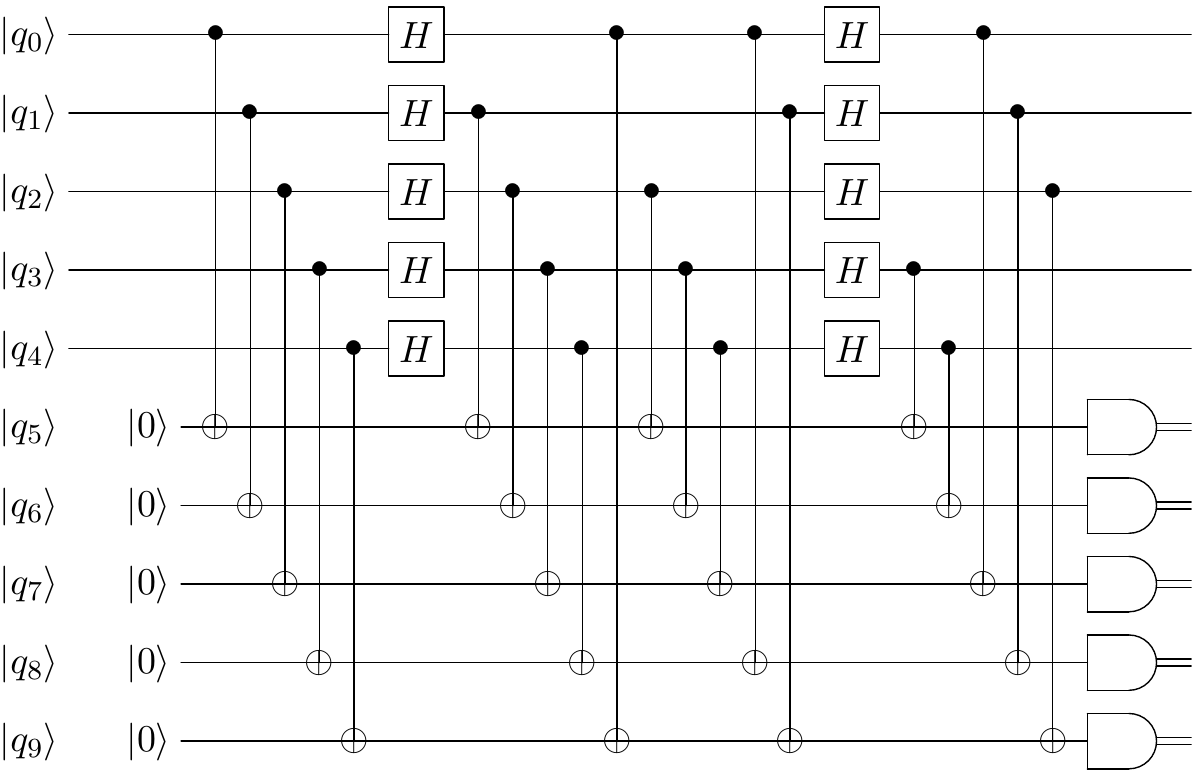}
  \caption{Cingle-cycle measurement circuit for the five-qubit code
    with $n_a=5$ ancillary qubits.}
  \label{fig:513-circ}
\end{figure}

To quote some numbers, Tables \ref{tab:rep-code} and
\ref{tab:rot-code} give the dimensions of generator matrices and
row-reduced generator matrices for two code families: the toy codes
$[[n_0,1,d_X=n_0/d_Z=1]]$ with stabilizer generators
$Z_iZ_{i+1\bmod n_0}$, $0\le i<n_0$, and the rotated toric codes (Ex.\
11 in Ref.~\onlinecite{Kovalev-Dumer-Pryadko-2011}) $[[n_0,1,2t+1]]$
with $n_0=t^2+(t+1)^2$ and stabilizer generators
$Z_i X_{i+t\bmod n_0}X_{i+t+1\bmod n_0}Z_{i+2t+1\bmod n_0}$, where
$0\le i<n_0$ and $t=1,2,\ldots$, with up to three measurements cycles.
The simplest examples of the measurement circuits used for these two
code families are shown in Figs.~\ref{fig:3-3-1-circ} and
\ref{fig:513-circ}, respectively.  In particular, while the original
generator matrix for the circuit EEG of the $[[13,1,5]]$ code with
measurements repeated $n_{\rm cyc}=3$ times has dimensions
$1066\times 1092$, row-reduction of rows of weight up to three reduces
the dimension to $182\times 247$.

\begin{table*}[htbp]
  \centering
  \begin{tabular}[c]{c|c|c|c||c|c|c|c|c||c|c||c}
    \hline
    \multicolumn{4}{c||}{circuit parameters}
    &\multicolumn{5}{c||}{generator matrix dimensions}
    &\multicolumn{2}{c||}{remaining columns $m_\epsilon$}&
    \\ \hline
    $n_0$&$n_a$&$n_{\rm cyc}$&$d_0$&orig&$w=2$&$w=3$&final&$w_{\rm fin}$&$\epsilon=10^{-2}$&$\epsilon=10^{-1}$&$\ell_1$\\
    \hline\hline
    3&3&1&3& $30\times36$&$3\times10$&$0\times10$&$0\times10$&0&10&10&7 
    \\                                                         
    5&5&1&5& $50\times60$&$5\times16$&$0\times16$&$0\times16$&0&16&16&11
    \\                                                         
    7&7&1&7& $70\times84$&$7\times22$&$0\times22$&$0\times22$&0&22&22&15
    \\ \hline 
    3&6&2&3& $72\times78$&$9\times19$&$3\times19$   &$2\times43$ &21&38&21&10
    \\                                                              
    5&10&2&5&$120\times130$&$15\times31$&$5\times31$&$3\times79$ &21&69&34&16
    \\                                                              
    7&14&2&7&$168\times182$&$21\times43$&$7\times43$&$4\times115$&21&99&48&22
    \\ \hline
    3&9&3&3&$102\times108$&$15\times28$&$6\times27$  &$4\times75$  &35&69&31&13\\
    5&15&3&5&$170\times180$&$25\times46$&$10\times45$&$7\times116$ &20&104&50&21\\
    7&21&3&7&$238\times252$&$35\times64$&$14\times63$&$10\times157$&20&140&69&29
    \\  \hline 
  \end{tabular}
  \caption{Parameters of the original and row-reduced generator
    matrices for repetition code circuits as in
    Fig.~\ref{fig:3-3-1-circ} but with $n_{\rm cyc}$ measurement
    cycles, $n_0$ data and $n_a=n_{\rm cyc}n_0$ ancillary qubits.
    Also shown are dimensions of row-reduced generator matrices with
    rows of weights $w=2$ and $w=3$ (and smaller) eliminated;
    $w_{\rm fin}$ is the minimum row-weight of the final generator
    matrix with the smallest number of rows remaining.  Remaining
    columns $m_\epsilon$ is the number of columns after columns with
    $|K_b|<\epsilon$ are dropped from the final generator matrix,
    assuming $p=0.05$ corresponding to a half-LLR value
    $K\approx 1.472$.  The last column gives the value of $\ell_1$ in
    the upper bound (\ref{eq:m-bound}) on the number of columns.}
  \label{tab:rep-code}
\end{table*}
\begin{table*}[htbp]
  \centering
  \begin{tabular}[c]{c|c|c|c||c|c|c|c|c|c||c|c|c||c}
    \hline
    \multicolumn{4}{c||}{circuit parameters}
    &\multicolumn{6}{c||}{generator matrix dimensions}
    &\multicolumn{3}{c||}{remaining columns $m_\epsilon$}&
    \\ \hline
    $n_0$&$n_a$&$n_{\rm cyc}$&$d_0$&orig&$w=2$&$w=3$&$w=4$&final&$w_{\rm fin}$
    &$\epsilon=10^{-3}$&$\epsilon=10^{-2}$&$\epsilon=10^{-1}$&$\ell_1$\\
    \hline\hline
    5&5&1&3&$130\times140$&$25\times35$&$20\times35$&$16\times43$&$13\times75$
                                            &10&65&47&23&11\\
    5&10&2&3&$280\times290$&$55\times65$&$45\times65$&$37\times81$&$32\times139$
                                            &10& 112&82&46&16\\
    5&15&3&3&$410\times420$& $85\times95$ & $70\times95$ &$58\times119$  &$51\times203$
                                            &10&159&118&70&21 \\ \hline
    13&13&1&5&$338\times364$&$65\times91$&$52\times91$&$40\times115$&$32\times163$
                                      &9&163&120&61&27\\ 
    13&26&2&5&$728\times754$&$143\times169$ &$117\times169$ & $93\times217$&$83\times291$
                                            &9&273 &213&118&40 \\ 
    13&39&3&5&$1066\times1092$&$221\times247$ &$182\times247$&$146\times319$&$134\times419$
                                            &9&384 &305 &182&53\\ \hline
    25&25&1&7&$650\times700$&$125\times175$&$100\times175$&$76\times223$ &$58\times331$
                                            &9&331 &234&116&51\\
    25&50&2&7&$1400\times1450$&$275\times325$ &$225\times325$ &$177\times421$ &$157\times555$
                                            &9&528 &413&223&76 
  \end{tabular}
  \caption{Same as in Tab.~\ref{tab:rep-code} but for rotated toric
    codes $[[t^2+(t+1)^2,1,2t+1]]$ with $t=1,2,3$, represented as
    single-generator cyclic codes, see Example 11 in
    Ref.~\onlinecite{Kovalev-Dumer-Pryadko-2011}.}
  \label{tab:rot-code}
\end{table*}

Second possibility for exact ML decoding is to perform a summation
over \emph{all} spins, as described in Sec.~\ref{sec:marginal-output}.
This gives a probability distribution directly for the error
equivalence classes; the logarithm of probability of an error
equivalent to $\mathbf{e}$ is given, up to an additive constant, by a
weighted sum of the half-LLR coefficients
$-2\sum_b e_bK_b^{\rm (fin)}$.  Unfortunately, such an exact
expression is expected to have an exponentially long list of
coefficients, see Eqs.~(\ref{eq:ell-one}) and (\ref{eq:m-bound}) for
an upper bound.  The corresponding column numbers are large even for
the relatively simple circuits in Tables \ref{tab:rep-code} and
\ref{tab:rot-code}.  In fact, the Mathematica program (which was not
written for efficiency) failed to complete full row-reduction except
for the simplest repetition codes with $n_{\rm cyc} =1$ round of
measurements.

In practice, the list of coefficients $K_b^{\rm (fin)}$ often contains
a large number of entries with small magnitudes.  This suggests a
range of approximations, where only sufficiently large coefficients
are preserved, e.g., $|K_b^{\rm (fin)}|>\epsilon$, for a given
$\epsilon>0$.  For incompletely reduced matrices in Tables
\ref{tab:rep-code} and \ref{tab:rot-code}, with $\epsilon=0.1$, the
number of columns is reduced, roughly, by a factor of two.  On general
grounds, the reduction factor is expected to be much larger for
fully-reduced generator matrices.

Alternatively, only a fixed number $\chi$ of the largest in magnitude
coefficients may be preserved.  This latter approach is similar
in spirit to approximate tensor network contraction using singular
value decomposition and a fixed maximum bond dimension.  Notice that
if only the coefficients corresponding to columns of the matrix
${\rm H}_0'$ in Eq.~(\ref{eq:block-H}) are preserved, we get an
approximation similar to the conventional phenomenological noise
model.

The ``history code'' decoding algorithm suggested by Chubb and
Flammia\cite{Chubb-Flammia-2018}, can be seen as a special case of
generator-based decoding.  Here the measurement circuit is assumed to
have a block structure, and summation is done over all circuit
locations with the exception of those at the output of each block.
With the circuits in Tables \ref{tab:rep-code} and \ref{tab:rot-code},
a block corresponds to a single measurement cycle.  The full
probability distribution is then recovered using Bayes rules, assuming
no correlations between measurements in different blocks.  Evidently,
even in this case, the complexity of the probability distribution
accounting for full error correlations could be prohibitive for exact
ML decoding.

\subsection{Parity-based decoding via marginal distributions}
\label{sec:dec-marg-H}
Summations over the spins in the subsets ${\cal I}_j$ with increasing
$j$ from a sequence like (\ref{eq:subset-sequence}) give marginal
distributions which account exactly for increasing numbers of
alternative spin configurations.  Respectively, the degeneracy of the
corresponding row-reduced half CSS codes decreases with increasing
$j$, down to a classical code with no degeneracy at the end of the
sequence, $j=s$.  A general expectation is that the accuracy of
minimum-energy (ME) decoding would be increasing with increasing $j$.
ME decoding becomes strictly equivalent to ML decoding for the
end-of-the-sequence classical code.

Formally, given a parity-check matrix $H$ and a set of LLR weights
$K_b$, ME decoding aims to find an error vector $\mathbf{e}$ which
gives the correst syndrome $\mathbf{e}H^T=\mathbf{s}$ and maximizes
the error likelihood $\sum_b (-1)^{e_b} K_b$ or, equivalently,
minimizes the error energy $\mathscr{E}=\sum_b e_b K_b$.  Compared to
generator-based decoding, an obvious advantage is that there is no
need to go over all $2^{2k}$ logical operators.  Unfortunately, for
generic codes, even the relatively simple problem of ME decoding has
an exponential complexity\cite{Iyer-Poulin-2013}.  Given that the
intermediate codes in the sequence (\ref{eq:subset-sequence}) tend to
be long, this makes it unpractical to use generic ME decoding
algorithms with exponential complexity, e.g., the information
subset\cite{Kruk-1989,Coffey-Goodman-1990} or the classical
Viterbi\cite{Viterbi-1967} algorithm.

Notice, however, that the sparsity of parity-check matrices
${\rm H}^{(j)}$ increase with $j$.  The final matrix ${\rm H}^{(s)}$
{is} expected to be sparse whenever the output code is an LDPC code.
Ideally, this classical code could be decoded with a linear
complexity using a variant of belief propagation (BP)
algorithm\cite{Gallager-book-1963,Fossorier-2001,%
  Richardson-Urbanke-2001,Declerq-Fossorier-Biglieri-2014}.  Assuming
a sufficiently small fraction of failed convergence cases, the result
would be equivalent to ML decoding of the correlated errors.

Unfortunately, this does not look so promising given the fact that
this final code is expected to have an exponential length, see
Eq.~(\ref{eq:m-bound}).  Additionally, as confirmed with limited
numerical simulations\cite{Zeng-Pryadko-2020}, having a large number
of small in magnitude LLR coefficients tends to reduce the convergence
rate.  Using approximate decoding schemes with reduced number of LLR
coefficients as discussed in Sec.~\ref{sec:dec-marg-G} is expected to
help with both issues.  Notice, however, that for such a reduction
certain columns in the generator matrix $\Theta$ are merely dropped
(puncturing).  The corresponding transformation for the parity-check
matrix ${\rm H}$ is \emph{shortening}\cite{MS-book}, which may reduce
the sparsity of the resulting matrix and, in turn, negatively affect
the convergence of BP decoding.

\section{Discussion}
\label{sec:discussion}

Improving the accuracy of syndrome-based decoding in the presence of
circuit-level error correlations would both increase the threshold to
scalable quantum computation and improve the finite-size performance
of quantum error correction.  Present results make two steps in this
direction.  First, the observation that ML decoding under these
conditions amounts to decoding the
code\cite{Bacon-Flammia-Harrow-Shi-2015,Bacon-Flammia-Harrow-Shi-2017}
associated with the circuit EEG, in the absence of correlations.
Second, the structure of this latter code can be significantly
simplified using row-reduction transformations, while leaving the
probability of ML decoding unchanged.  A variety of approximate
decoding schemes naturally follow, which interpolate between the exact
ML decoding and the decoding within a relatively simple
phenomenological error model, with an additional handle on the
degeneracy of the actual code to be used for decoding.  Designing
practical decoding algorithms, e.g., in application to surface codes
with well-developed near-optimal syndrome measurement circuits, would
require a substantial additional effort.  However, this does not look
like an unsolvable problem.

{Decoding quantum LDPC codes} is a major problem in the theory of
QECCs, especially in a fault-tolerant regime with syndrome measurement
errors present.  While a substantial progress has been made in recent
years\cite{Fawzi-Grospellier-Leverrier-2017,%
  FGL18,Grospellier-Krishna-2018,Panteleev-Kalachev-2019,%
  Grospellier-Groues-Krishna-Leverrier-2020}, this
problem remains open in application to generic highly-degenerate
codes.  Transformations discussed in Sec.~\ref{sec:marginal} change
the degeneracy of a quantum code, and can even map from a quantum to a
classical code.  This opens new avenues to explore in decoding, in
particular, new ways to apply existing iterative decoding algorithms
to highly degenerate codes.

{Circuit optimization:} Traditional approach to quantum error
correction is to start with a code, come up with an FT measurement
circuit, compile it to a set of gates available on a specific quantum
computer, and then finally design a decoder.  Instead, one could start
with the list of permitted two-qubit gates on a particular device and
enumerate all good error-detecting circuits, increasing circuit depth
and the number of gates.  Given the circuit, it is easy to find the
parameters of the input/output codes, as well as construct the
associated circuit code.  While at the end we would still need to
evaluate and compare the performance of thus constructed codes, such a
procedure could offer a shortcut to circuit optimization for specific
hardware.

\textbf{Acknowledgment:} This work was supported in part by the NSF
Division of Physics via grant No.\ 1820939.

\appendix

\section{Ranks of the matrices}
\label{sec:app-circ}

\subsection{Rank of the circuit-EEG generator matrix}
\label{sec:app-rank-G}
Consider a good error-detecting circuit in Fig.~\ref{fig:meas-circ}
with the constant $c=1/2^\kappa$ in Eq.~(\ref{eq:projector}) and the
input (or output) code encoding $k=n_0-r_0$ qubits, where $r_0$ is the
rank of the input-code stabilizer group.  Here $\kappa$ is an
integer\cite{Bacon-Flammia-Harrow-Shi-2017}.  Also assume that $f$
measurements are redundant, so that the total number of ancillary
qubits is $n_a=\kappa+r_0+f$.  Generators of the circuit EEG can be
used to propagate any circuit error all the way to the output layer on
the right, which requires $2N-2(n_0+n_a)$ independent generators,
where $N$ is the number of circuit locations.  In addition, there are
$n_a$ ancillary $Z_j$ generators on the input and $n_a$ on the output
layers.  However, not all of them are independent.  Indeed, when the
$n_a$ ancillary $Z_j$ operators are propagated to the right, we get
$r_0$ independent operators, each containing a product of ancillary
$Z_j$ and an element of the stabilizer group, $\kappa$ independent
operators containing ancillary $X_j$, and $f$ additional combinations
that are redundant except for a product of ancillary $Z_j$.  Overall,
this gives $\rank G=2N-2(n_0+n_a)+n_a+\kappa+r_0$, which is the same
as in Eq.~(\ref{eq:rank-G}).

\subsection{Rank of the circuit stabilizer group}
The circuit stabilizer group must commute with any EEG generator and
also with any logical operator of the output code (say).  Necessarily,
its elements must be spackles\cite{Bacon-Flammia-Harrow-Shi-2017}.
Any spackle is uniquely determined by its support on the input layer,
thus, there are total of $2(n_0+n_a)$ independent spackles.  We also
have to ensure commutativity with ancillary $Z_j$ generators on the
left (drop $n_a$ spackles) and on the right (drop $\kappa$ spackles).
Additional $r_0$ spackles have to be removed to ensure commutativity
with the elements of the output code stabilizer generators, and $2k$
spackles to ensure commutativity with the output code logical
operators.  Overall, this leaves
\begin{eqnarray*}
  \rank H&=&2(n_0+n_a)-n_a-\kappa-r_0-2(n_0-r_0)\\
         &=&n_a-\kappa+r_0,
\end{eqnarray*}
which is exactly the result in  Eq.~(\ref{eq:rank-H}).

\bibliography{qc_all,more_qc,lpp,neural,spin,percol,sg,ldpc}
\end{document}